\documentclass[twoside]{sfm}

\input{sf.def}

\begin{document}

\def\llm{{\sc LLmodels}}
\def\atl{{\sc ATLAS9}}
\def\aatl{{\sc ATLAS12}}
\def\starsp{{\sc STARSP}}
\def\aur{$\Theta$~Aur}
\def\logg{\log g}
\def\tauros{\tau_{\rm Ross}}
\def\kms{km\,s$^{-1}$}
\def\bz{$\langle B_{\rm z} \rangle$}
\def\degr{^\circ}
\def\aaps{A\&AS}
\def\aap{A\&A}
\def\apjs{ApJS}
\def\apj{ApJ}
\def\rmxaa{Rev. Mexicana Astron. Astrofis.}
\def\mnras{MNRAS}
\def\actaa{Acta Astron.}
\newcommand{\Tef}{T$_{\rm eff}$~}
\newcommand{\Vt}{$V_t$}
\newcommand{\CC}{$^{12}$C/$^{13}$C~}
\newcommand{\CDC}{$^{12}$C/$^{13}$C~}

\pagebreak

\thispagestyle{titlehead}

\setcounter{section}{0}
\setcounter{figure}{0}
\setcounter{table}{0}

\markboth{Hubrig et al.}{O and B-type stars}

\titl{HARPS spectropolarimetry of O and B-type stars}{Hubrig S.$^1$, Sch\"oller M.$^2$, Ilyin I.$^1$, Lo Curto G.$^2$}
{
 $^1$Leibniz-Institut f\"ur Astrophysik, An der Sternwarte~16, 14482~Potsdam, Germany, email: {\tt shubrig@aip.de} \\
 $^2$European Southern Observatory, Karl-Schwarzschild-Str.~2, 85748~Garching, Germany
}

\abstre{
Our knowledge of the presence and the strength of magnetic fields in massive 
O and B-type stars remains very poor.
 Recent observations indicate that the presence of magnetic 
fields is responsible for a wide range of phenomena observed in massive stars at different wavelengths,
such as chemical 
peculiarity, excess of emission in UV-wind lines and periodic UV wind-line variability, unusual 
X-ray emission, and  cyclic 
variability 
in H$\alpha$ and He~II 
$\lambda$4686.
However, it is difficult to establish relationships to multiwavelength diagnostics, as the strength of the 
detected magnetic fields and their geometry
differ from one star to the other.
In this work, we present new magnetic field measurements in a number of O and B-type 
stars of different classification observed with HARPS in spectropolarimetric mode.
}

\baselineskip 12pt

\section{Introduction}
\label{sect:hubrig_intro}

Recent observational and theoretical results emphasised the potential significance of magnetic fields for stellar structure, 
evolution, and circumstellar environment of massive O- and B-type stars.  Our previous searches for magnetic fields in massive 
stars were mostly based on FORS\,1/2 low-resolution polarimetric spectra obtained at ESO/Paranal. On the other hand, a large 
number of spectropolarimetric HARPSpol observations became publically available in the ESO archive in recent years, among 
them the spectra of pulsating $\beta$\,Cephei stars, Be stars, and massive O-type stars. In this work, we present new magnetic field 
measurements for the two $\beta$\,Cephei stars HD\,55857 and HD\,64503 with spectral types B0.5V and  B2.5V, respectively; the Be star 
HD\,92939 with spectral type B4V, the blue straggler $\theta$\,Car, and two O-type stars, $\zeta$\,Pup with spectral type O4f(n)p, 
and the 
Of?p star CPD~$-$28\,2561. Among this sample, results of magnetic field measurements were previously published only for $\theta$\,Car 
and the Of?p star CPD~$-$28\,2561 (Hubrig et al.\ 2008 \cite{hubrig_Hubrig2008}; 2011a \cite{hubrig_Hubrig2011a}; 2013 \cite{hubrig_Hubrig2013}).

\section{Magnetic field measurements in B-type stars}
\label{sect:hubrig_bstars}

\begin{figure}[!t]
\begin{center}
\includegraphics[width=0.45\textwidth, angle=0]{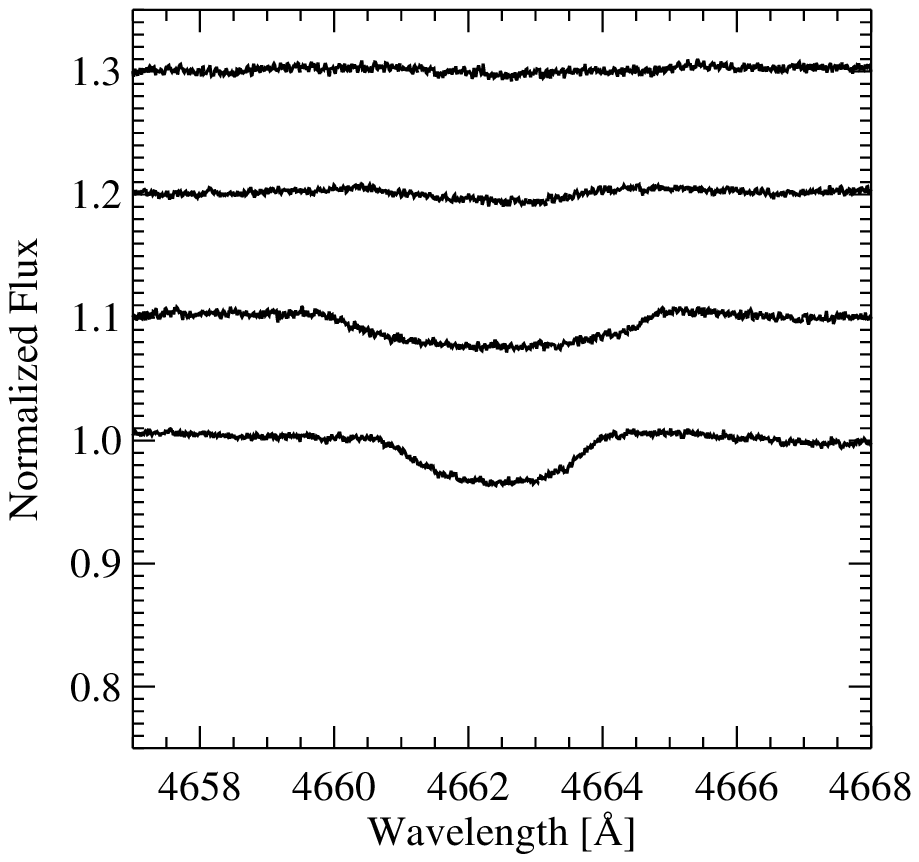}
\includegraphics[width=0.45\textwidth, angle=0]{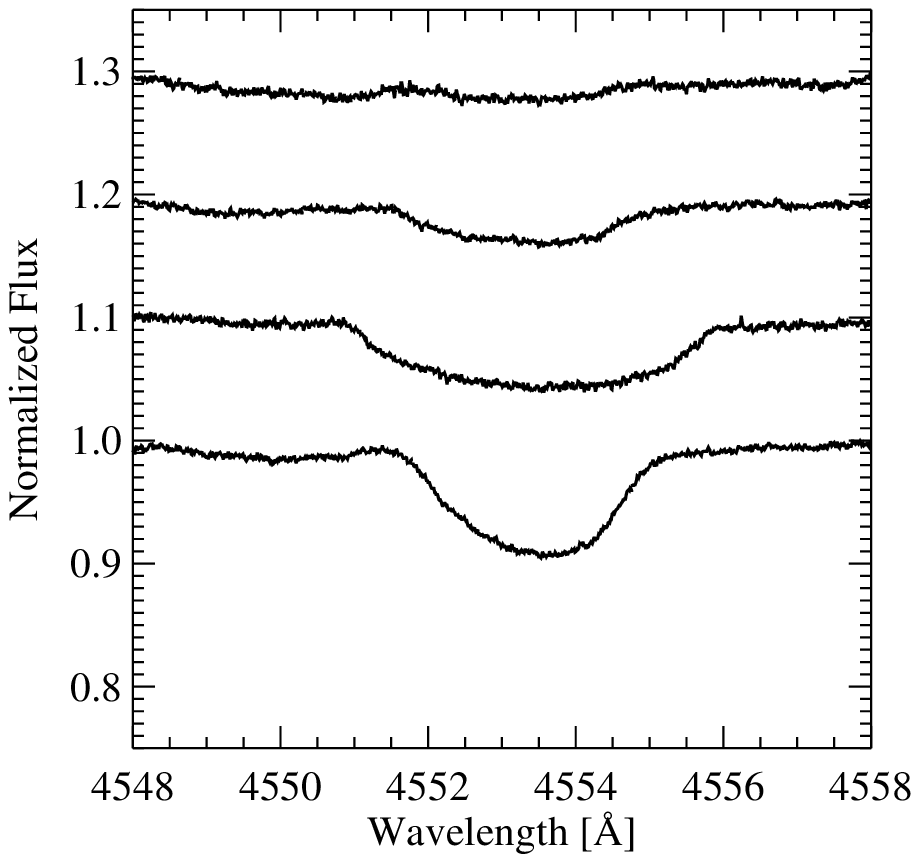}
\vspace{-5mm}
\caption[]{
{\em Left:} HARPSpol Stokes~$I$ spectra (from top to bottom) of HD\,92238, HD\,55857, HD\,64503, and $\theta$\,Car in the 
region containing the O\,II\,$\lambda$4662 line.
{\em Right:} HARPSpol Stokes~$I$ spectra (from top to bottom) of HD\,92238, HD\,55857, HD\,64503, and $\theta$\,Car in the 
region containing the Si\,III\,$\lambda$4552 line.
}
\label{fig:hubrig_cep_o_si}
\end{center}
\end{figure}

For the measurements of the magnetic 
fields in the sample of stars observed with HARPSpol, we used the moment technique developed by Mathys (e.g.\
Mathys 1991 \cite{hubrig_Mathys1991}).  
All four studied B-type stars have large $v\,\sin\,i$ values, from 113\,km/s for $\theta$\,Car up to 187\,km/s for HD\,64503 (Telting et al.\
2006 \cite{hubrig_Telting2006}).
In Fig.~\ref{fig:hubrig_cep_o_si}, from top to bottom, we present the spectra of HD\,92238, HD\,55857, HD\,64503, and 
$\theta$\,Car in regions around the O\,II\,$\lambda$4662 and Si\,III\,$\lambda$4552 lines. For the magnetic field measurements, we used 
all unblended metal lines, and He\,I lines. The He\,II\,$\lambda$4686 line was used only in the measurements of the two hottest stars, 
$\theta$\,Car and HD\,55857. 

\begin{figure}[!t]
\begin{center}
\includegraphics[width=0.45\textwidth, angle=0]{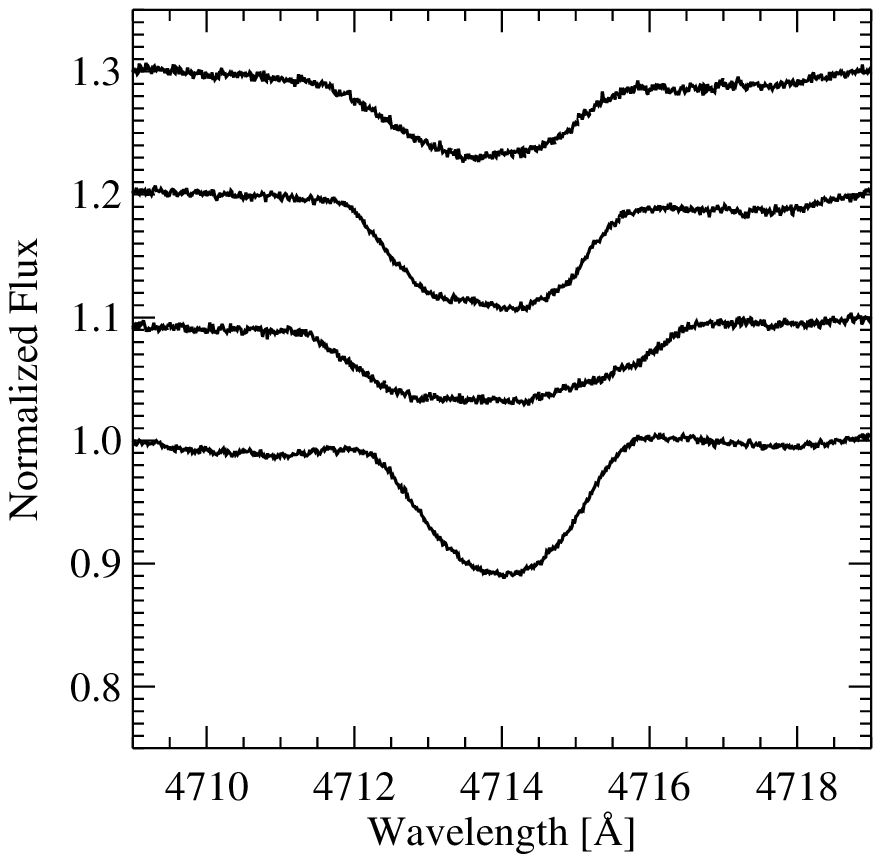}
\includegraphics[width=0.45\textwidth, angle=0]{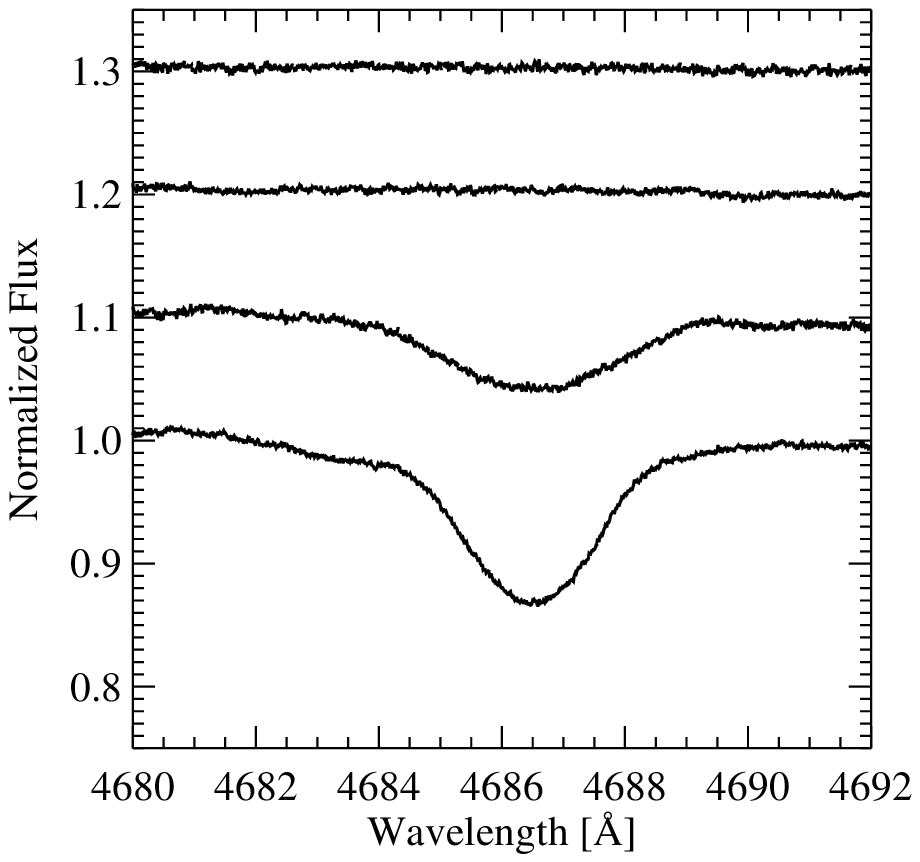}
\vspace{-5mm}
\caption[]{
{\em Left:} HARPSpol Stokes $I$ spectra (from top to bottom) of HD\,92238, HD\,55857, HD\,64503, and $\theta$\,Car in the 
region containing the He~I\,$\lambda$4713 line.
{\em Right:} HARPSpol Stokes $I$ spectra (from top to bottom) of HD\,92238, HD\,55857, HD\,64503, and $\theta$\,Car in the 
region containing the He~II\,$\lambda$4686 line.
}
\label{fig:hubrig_cep_he}
\end{center}
\end{figure}

The spectra of the same stars in the spectral regions around the He\,I\,$\lambda$4713 and He\,II\,$\lambda$4686 lines are presented in Fig.~\ref{fig:hubrig_cep_he}.
The most recent abundance analysis of $\theta$\,Car showed  nitrogen overabundance typical of the values found for other slowly-rotating 
(magnetic) B-type dwarfs and carbon depletion (Hubrig et al.\ 2008 
\cite{hubrig_Hubrig2008}). For this star, we attribute the chemical peculiarities to a past episode of mass transfer between the two binary components. 
Among 26 magnetic field measurements using FORS\,1 spectropolarimetric data, only a few measurements had a significance level 
of 3$\sigma$. On the other hand, a periodicity of the order of 8.8\,min was detected in the magnetic data.
Such a  periodicity
is surprising, and if it is not spurious, its discovery would give
rise to the important question whether the presence of pulsations
could cause such a periodicity. No studies of short-time pulsations
exist for $\theta$~Car so far. B0 main-sequence stars are expected
to pulsate in low radial order p- and g-modes with periods of the
order of hours (typically 3 to 8\,h). For such massive stars, periods
of the order of a few minutes would correspond to high radial
order p-modes, but non-adiabatic codes do not predict their excitation.
However, since current models do not take into account
the presence of a magnetic field, one cannot exclude the possibility
that magnetic fields would favour the excitation of these
types of modes in analogy to roAp stars, which do possess a magnetic
field and pulsate in high radial order modes of the order of
a few minutes.

The results of our measurements are 
presented in Table~\ref{tab:hubrig_mf_HARPS}. The first column lists the name of the target, followed by the MJD value, 
signal-to-noise ratio of the studied spectrum, and the measured longitudinal magnetic field. In spite of the very high S/N achieved in the HARPSpol observations, 
the accuracy of the measurements in the rather fast rotating 
massive stars is relatively low.
Formally significant detections above the 3$\sigma$ level were achieved in the $\beta$\,Cephei stars 
HD\,58647 and HD\,64503. 

\begin{table}
\centering
\caption{Measurements of the mean longitudinal magnetic field using high-resolution HARPS spectra.}
\label{tab:hubrig_mf_HARPS}
\tabcolsep1.2mm
\begin{tabular}{lrrr@{$\pm$}r}
\hline
\hline
\multicolumn{1}{c}{Object} &
\multicolumn{1}{c}{MJD} &
\multicolumn{1}{c}{S/N} &
\multicolumn{2}{c}{$\left<B_{\rm z}\right>$} \\
\hline
HD\,58647 & 55913.183 & 560 & $-$212 & 64 \\
HD\,64503 & 55912.279 & 660 &    127 & 38 \\
HD\,92938 & 55707.959 & 560 & $-$117 & 48 \\
$\theta$\,Car & 55706.095 & 710 & 72 & 48 \\
$\zeta$\,Pup & 55910.226 & 600 & 145 & 93 \\
\hline
\end{tabular}
\end{table}

The obtained values of longitudinal magnetic fields listed in Table~\ref{tab:hubrig_mf_HARPS}
clearly confirm our previous conclusions that magnetic fields of kG order cannot be expected 
in $\beta$\,Cephei and Be stars (Hubrig et al.\ 2006 \cite{hubrig_Hubrig2006}; 2009a \cite{hubrig_Hubrig2009a}; 2009b \cite{hubrig_Hubrig2009b}).

\section{Magnetic field  measurements of the O4 star $\zeta$\,Pup and the Of?p star CPD~$-$28\,2561}
\label{sect:hubrig_ostars}

$\zeta$\,Pup, which is one of the closest and brightest massive stars, is a fast rotating ($v\,\sin\,i$ is about 200\,km/s) runaway star,
having had a possible encounter with the cluster Trumpler~10 (Hoogerwerf et al.\ 2001 \cite{hubrig_Hoogerwerf2001}). 
No magnetic field is detected in this star, taking note that the measurement accuracy 
is the lowest among the sample stars. 

A rather strong magnetic field in the Of?p star CPD~$-$28\,2561, of the order of a few hundred Gauss, was detected a few years 
ago by Hubrig et al.\ (2011a \cite{hubrig_Hubrig2011a}; 2013 \cite{hubrig_Hubrig2013}) using FORS\,2 observations. 
Walborn (1973 \cite{hubrig_Walborn1973}) introduced the Of?p category for massive O stars displaying recurrent spectral variations in 
certain spectral lines, sharp emission or P Cygni profiles in He~I and the Balmer lines, and strong C~III emission lines around $\lambda$4650.
Spectropolarimetric observations on three consecutive nights in 2011 revealed strong variations in both the longitudinal magnetic field strength 
and several hydrogen and helium line profiles (Hubrig et al.\ 2012 \cite{hubrig_Hubrig2012}).
Unfortunately, 
the HARPSpol spectrum of this star has very 
low S/N, only about 100, preventing accurate magnetic field measurements. Our measurements indicate that the field has negative
polarity and is at least 400--500\,G strong.

\section{Summary}
\label{sect:hubrig_summary}

Our previous magnetic field measurements using the VLT instruments FORS\,1/2 in spectropolarimetric mode revealed the presence 
of weak magnetic fields in a few $\beta$\,Cephei stars, among them the record holder $\xi^1$\,CMa with a magnetic field of the 
order of 300--400\,G (Hubrig et al.\ 2006 \cite{hubrig_Hubrig2006}; 2009 \cite{hubrig_Hubrig2009a}; 2011b \cite{hubrig_Hubrig2011b}).
Recent high-resolution 
spectropolarimetric observations of two stars of this type with HARPSpol support their weak 
magnetic nature with detections at a significance level of about 3$\sigma$. Future magnetic field measurements are urgently needed 
to determine the role of weak magnetic fields in modeling oscillations of B-type stars.


%
%
%
%

\end{document}